\documentclass[12pt,preprint]{aastex}

\slugcomment{Not to appear in Nonlearned J., 45.}

\shorttitle{AGN-HOST CONNECTION IN HYBRID QSOs}
\shortauthors{Wang \& Wei}

\begin{document}


\title{UNDERSTANDING THE AGN-HOST CONNECTION IN BROAD \ion{Mg}{2} EMISSION SELECTED
AGN-HOST HYBRID QAUSARS 
}

\author{J. Wang\altaffilmark{1} and J. Y. Wei\altaffilmark{1}}
\affil{National Astronomical Observatories, Chinese Academy of Science, 
20A Datun Road, Chaoyang District, Beijing 100012, China}
\email{wj@bao.ac.cn}



\begin{abstract}

We study the issue of AGN-host connection in intermediate-z ($1.2>z>0.4$) galaxies 
with hybrid spectra (hybrid QSOs for short). 
The observed spectra redward of the Balmer limit are dominated by starlight, and 
the spectra at blue end by both an AGN continuum and a \ion{Mg}{2} broad emission line. 
Such unique property 
allows us to examine both AGN and its host galaxy in individual galaxy simultaneously.
At first, 15 hybrid QSOs are selected from the Sloan Digital Sky Survey
Data Release 6. The spectra are then analyzed in detail in three objects: 
SDSS\,J162446.49+461946.7, SDSS\,J102633.32+103443.8 and SDSS\,J090036.44+381353.0. 
Our spectral analyzing shows that the current star formation activities are strongly
suppressed, and that the latest burst ages range from $\sim$400Myr to 1Gyr.
Basing upon the \ion{Mg}{2}-based black hole masses, the three hybrid QSOs
are consistent with the $D_n(4000)-L/L_{\mathrm{Edd}}$ sequence that was previously
established in local AGNs. The three hybrid QSOs are located in the middle range 
of the sequence, which implies that the hybrid QSOs are at the transition 
stage not only from young to old AGN, but also from host-dominated phase to AGN-dominated phase. 

\end{abstract}





\section{INTRODUCTION}

Despite the fact that there are still a great deal of unresolved problems, 
recent numerous observational and theoretical studies showed that 
Active Galactic Nucleus (AGN) plays an important role in galaxy evolution.
The mass of 
central supermassive black hole (SMBH) is believed to growth simultaneously with 
the formation of the bulge of the host where the SMBH resides in as clued 
by the well-established $M_{\mathrm{BH}}-\sigma_*$ relation (e.g.,
Magorrian et al. 1998; Gebhardt et al. 2000; Tremaine et al. 2002; 
Ferrarese \& Merritt 2000; Ferrarese et al. 2006; Greene \& Ho 2006; Ho et al. 2008a,b).    
Another clue of the coevolution of SMBH and starburst is the fact that 
both quasar activity density and star formation rate (SFR) density apparently peak at 
$z\approx 2-3$ (e.g., Nandra et al. 2005). By adopting a variety of 
mechanisms of feedback from black hole growth on star formation 
occurring in the bulge, a series of models were developed
to explain the co-evolution of AGN and its host galaxy (e.g., Fabian 1999; 
Begelman \& Nath 2005; Granato et al. 2004; Springel et al. 2005a,b; Hopkins et al. 2006 
and references therein).

Understanding how AGN co-evolves with star formation in detail is, however,
seriously impeded by the trouble caused by the orientation effect (see review in Antonucci 1993;
Elitzur 2007). Briefly, the starlight component is usually masked by
the strong continuum and broad emission lines emitted from the central engine
in the spectra of type I AGNs. On the contrary, the 
starlight component can be much more easily detected in type II AGNs, 
however the AGN continuum and broad
lines are blocked by the torus. So far, a variety of 
approaches have been adopted to overcome the problem by different authors. 
We refer the readers to Wang \& Wei (2008, and references therein) 
for a recent comment on these approaches. 
In Wang \& Wei (2008) and Xiao et al. (in preparation), these authors examined
the co-evolution of AGN and star formation by using the local partially
obscured AGNs (i.e., Seyfert 1.8/1.9 galaxies) selected from the MPA/JHU SDSS DR4
catalog (e.g., Kauffmann et al. 2003a,b,c). The spectra of these galaxies show 
good balance between the AGN and starlight components, which allows the authors to directly
determine the properties of both AGN and
stellar population simultaneously in individual object. They then proposed an 
evolutionary scenario that the Eddington ratio decreases as 
the circumnuclear stellar population ages (see also in Heckman et al. 2004; 
Kewley et al. 2006, Wild et al. 2007).

We attempt to extend these studies to objects within $1.2>z>0.4$ in this paper. 
Since the emission of evolved stars drops significantly at short wavelength region,
the broad \ion{Mg}{2}$\lambda$2800 emission line could be used to directly estimate the 
properties of central SMBH when broad H$\beta$ emission is weak or strongly contaminated by 
the star light and when H$\alpha$ emission is shifted out of the optical region. 
Here, we focus on the 
quasars with hybrid spectral properties (hereafter ``hybrid QSOs'' for short): 
a) the spectrum redward of the Balmer limit is dominated by a starlight component; 
b) at the same time, the spectrum at blue end shows an evident broad \ion{Mg}{2}$\lambda$2800 
emission line that is emitted from the central AGN.
To our knowledge at present, similar object is only
studied in detail (for stellar population only) in the particular post-starburst QSO, 
UN\,J1025-0040, by Brotherton et al. (1999). Another striking case with similar spectrum is reported in
SDSS\,J231055-090107 (Canalizo et al. 2006).

In this paper, we first select a sample of 15 hybrid QSOs from the SDSS. 
The detailed spectral analysis is then performed in three objects: SDSS\,J163446.49+461946.7
(at $z=0.576$, abbreviated here as SDSS\,J1634+4619), SDSS\,J102633.32+103443.8 (at $z=0.435$, 
abbreviated here as 
SDSS\,J1026+1034) and SDSS\,090036.44+381353.0 (at $z=0.434$, abbreviated here as SDSS\,0900+3813).
The spectra of the three objects have high adequate average signal-to-noise (S/N) ratios of whole spectra (see Table 1),
appropriate \ion{Mg}{2} emission profiles and stellar absorption features (especially the 4000\AA\ break 
features).  
The sample selection and spectral analysis are described
in \S 2 and \S 3, respectively. \S 4 presents the results and discussions.
The $\Lambda$CDM cosmology with parameters 
$h_0=0.7$, $\Omega_M=0.3$ and $\Omega_\Lambda=0.7$ (Spergel et al. 2003) is adopted 
throughout the paper.

\section{SAMPLE SELECTION}

SDSS is an ambitious project designed to eventually survey 
one-quarter of the entire sky in images and spectra (York et al. 2000). The 
survey is carried out by a dedicated wide-field (3\symbol{23}) 2.5m telescope
located at Apache Point Observatory. The telescope is equipped with two fiber-fed
spectrographs and a mosaic CCD camera. Each spectrum is taken with a 
3\symbol{125} diameter fiber aperture. The spectra have a 
resolution $R\sim1800$ (corresponds to $\sigma_{\mathrm{int}}\sim65\ \mathrm{km\ s^{-1}}$), 
and cover a wavelength range 
from 3800 to 9200\AA\ in the observer's frame. The spectra are spectrophotometrically
calibrated within an accuracy of $\sim$20\% by observing subdwarf F stars in each 
3\symbol{23} field of view. The raw spectra are reduced automatically by the developed 
pipelines, spectro2d and specBS (Glazebrook et al. 1998; Bromley et al. 1998).

At first, according to the redshifts determined by the SDSS pipelines, 
we select the galaxies and quasars within $1.2>z>0.4$ from the SDSS 
Data Release 6 catalogs (Adelman-McCarthy et al. 2008). The redshift range 
ensures that both \ion{Mg}{2}$\lambda$2800 emission line 
and 4000\AA\ feature (at the rest frame) shift into the wavelength coverage at observer's frame. 
The \ion{Mg}{2}$\lambda$2800 line emission
is then required to be detected above 3$\sigma$ significance level according to
the emission-line measurements provided by the SDSS pipelines.
The selected sample contains 1,946 galaxies and 7,694 quasars fulfilling the 
above criteria. An automatic routine is developed to select the hybrid quasars
whose spectra have an evident break around the rest-frame wavelength 4000\AA,
i.e., $D_n(4000)>1$, where 
$D_n(4000)=\int_{4000}^{4100}f_\lambda d\lambda/\int_{3850}^{3950}f_\lambda d\lambda$ 
(Bruzual 1983; Balogh et al. 1999). The selected spectra of quasars and galaxies are then 
inspected by eyes one by one. Finally, the selection results in a sample of 
totally 15 candidates of hybrid QSOs. We emphasize that the selection outlined above
provides a representative sample of hybrid QSOs (though not necessarily complete).
Given the $D_n(4000)$ criterion, our selection would miss the QSOs whose spectra are
extremely blue or dominated by extremely hot stars. 
In the first case, the blue AGN continuum dilutes the 4000\AA\ drop that is caused by the 
stellar absorptions. The starlight component is therefore hard to be identified and separated
from the observed spectra (the same for the second case). The sample is listed in Table 1. 
For each object, the table shows its plate number, modified Julian date of the 
observation, number of fiber on the plate, redshift, equivalent width 
of the \ion{Mg}{2} emission line, corresponding S/N ratio of \ion{Mg}{2}, and average 
S/N ratio of the whole spectrum. The observed spectra 
are displayed in Figure 1.

Except the three objects
SDSS\,J1634+4619, SDSS\,J1026+1034 and SDSS\,0900+3813, detailed spectral analysis 
is a hard task for most of the candidates listed in the sample at the current stage, either 
because of the low average S/N ratios of the whole spectra or because of the poor \ion{Mg}{2}
profiles and blurry 4000\AA\ break features. In order to determine their physical properties,
deeper spectroscopic observations are necessary for the remaining objects. 
At first glance each of the three spectra
is dominated by both an underlying powerlaw continuum and a 
broad \ion{Mg}{2} emission line attributed to AGN at the blue part, and by a
spectrum of an intermediate-massive star (i.e., A to G type) at the red part. 
The evident starlight component allows us to 
separate the contribution
of the host galaxy from each observed spectrum. Additionally, the properties of the black hole accretion 
could be derived from the evident \ion{Mg}{2} broad emission line.

\section{SPECTRAL ANALYSIS}

The observed spectra are at first smoothed with a boxcar of 6 pixels 
($\sim$10\AA) to enhance the S/N ratios and the accuracy of spectral
measurements. The smoothed spectra are then reduced through the standard procedures by 
IRAF package\footnote{IRAF is distributed by the National
Optical Astronomical Observatories, which is operated by the Association of Universities
for Research in Astronomy, Inc., under cooperative agreement with the 
National Science Foundation.}, including Galactic extinction and redshift 
corrections. The Galactic extinction is corrected for each spectrum 
by the color excess 
$E(B-V)$ adopted from NED\footnote{The Schlegel, Finkeiner, and Davis Galactic reddening maps (Schlegel et al. 1998)
are adopted for NED.} assuming an extinction curve with $R_V=3.1$ (Cardilli et al. 1989).  
Each Galactic extinction-corrected spectrum is then transformed to the 
rest frame according to the redshift given by the SDSS pipelines.
The reduced spectra at the rest frame are displayed in Figure 2 (see the solid curve in each panel).
Each spectrum clearly shows not only a discontinuity at around 4000\AA\ caused by the metal absorptions of stars, 
but also a strong 
\ion{Mg}{2} broad emission line superposed on an underlying AGN continuum at blue end.
Evident H$\delta$ and \ion{Ca}{2} H,K absorption features can be identified in 
the spectra of both SDSS\,J1634+4619 and SDSS\,J1026+1034. Although the spectrum of SDSS\,0900+3813 shows 
relatively weak H$\delta$ absorption, the \ion{Ca}{2} H,K absorptions are still 
strong. The presence of nuclear accretion activity in the three objects is 
additionally supported by the marginally detectable high ionization lines, 
such as [\ion{Ne}{3}]$\lambda$3868, [\ion{Ne}{5}]$\lambda\lambda$3426, 3346.

The single stellar population (SSP) models developed by Bruzual \& Charlot (2003, BC03) 
are used to interpret the observed starlight spectra for the three hybrid QSOs. 
At the beginning,
we extract a series of spectra from the BC03 SSP models with the Chabrier IMF at different 
metallicities. For a given metallicity, we attempt to reproduce each observed 
starlight component by the linear combination of five 
instantaneous bursts at different ages. The ages range from 0.4 to 4.0Gyr (i.e.,
0.4, 0.6, 0.8, 1.0, 4.0Gyr). Since a starlight spectrum is predicted to evolve slightly after 1.0 Gyr, 
the 4.0Gyr old stellar population is used to reproduce the underlying old stellar 
population. The bursts younger than 0.4Gyr are not considered in our spectral modeling
both because of the evident H$\delta$ and \ion{Ca}{2} absorptions and 
because of the likely degeneracy between the spectrum of very young stellar population
and AGN blue continuum.    

In summary, in order to model the observed spectra, our template contains the
starlight spectra of the five instantaneous bursts at the given stellar population ages, 
a powerlaw continuum and an UV \ion{Fe}{2} template 
that are both attributed to AGNs, and a Galactic extinction curve (Cardelli et al. 1989).   
Bruhweiler \& Verner (2008) recently calculated a grid of \ion{Fe}{2} emission
spectra. The predicted spectrum giving the best fit to the observed I\,ZW\,1 spectrum
is adopted in the current study. The adopted \ion{Fe}{2} template is calculated
for $\log[n_{\mathrm{H}}/(\mathrm{cm^{-3}})]=11.0$, 
$\log[\Phi_{\mathrm{H}}/(\mathrm{cm^{-2}\ s^{-1}})]=20.5$, $\xi/(1\ \mathrm{km\ s^{-1}})$, and
830 energy levels. The template is broadened by convolution with a Gaussian profile having
the same width as the \ion{Mg}{2} broad emission before our spectral modeling.

A $\chi^2$ minimization is performed over the rest wavelength
range from 2900\AA\ to 5500\AA, except the range around the 
strong emission lines (e.g., \ion{Mg}{2}$\lambda2800$, H$\beta$, [\ion{O}{3}]$\lambda$5007
and [\ion{O}{2}]$\lambda$3727). For each object, the fitting is carried
out for the several different metallicities (from 0.02$Z_\odot$ to 2.5$Z_\odot$)
to test if the results are robust in light of
changing the metallicity. The fitting at extremely low metallicity (i.e., 0.02$Z_\odot$) is 
excluded in the subsequent studies for SDSS\,0900+3813 because
a relatively 
high metallicity is required to reproduce the observed metal absorption lines.    
The fittings at the solar metallicity 
(the base models, see below) are illustrated in Figure 2. 
Figure 3 illustrates the fraction of mass associated to each of the five instantaneous bursts
used in our fitting. As shown in the figure, the modeling results at the different metallicities 
are highly consistent with each other. For SDSS\,J1634+4619 and SDSS\,J1026+1034,
their spectra can be interpreted by the combination of an old stellar population ($>1$Gyr) 
and a younger one (a few of 100Myr). Although there are uncertainties in the precise age
estimates due to both metallicity effect and used models, a $\leq400$Myr and a $\sim600$Myr old
young stellar populations are specifically required in SDSS\,J1026+1034 and SDSS\,J1634+4619, respectively. 
In contrast, SDSS\,0900+3813 is dominated by a 
relatively old stellar population ($\sim1$Gyr), which is in agreement with its relatively
weak H$\delta$ absorption feature that is already mentioned above.  
Since the spectral modeling results change little when the different metallicities are considered, 
the models with the solar metallicity are then used as the base models in our subsequent studies.

\section{RESULTS AND DISCUSSION}

Table 2 lists the results of the spectral measurements that are performed in the rest frame.
The residual emission-line spectra are used to measure emission line features, and the modeled 
starlight spectra to measure the Lick indices. For each object, 
lines (1) and (2) lists the redshift and modeled 
intrinsic extinction, respectively.  
Lines (2) to (5) list the AGN continuum flux at the rest-frame 
wavelength 3000\AA, the [\ion{O}{3}] and [\ion{O}{2}] emission line fluxes, and the FWHM of 
the \ion{Mg}{2} broad emission line. All the fluxes quoted above are corrected for the 
modeled intrinsic extinction. The fluxes of [\ion{O}{3}] and [\ion{O}{2}] emission line are
measured by direct integration\footnote{
The reported statistic errors are
derived by the method given in Perez-Montero \& Diaz (2003). That is $\sigma_{\mathrm{l}}=
\sigma_c N^{1/2}[1+EW/(N\Delta)]^{1/2}$, where $\sigma_l$ is the error of line flux, $\sigma_c$
the standard deviation of continuum in a box near the line, $N$ the number of pixel used to measure
the line flux, $EW$ equivalent width of the line, $\Delta$ the wavelength dispersion in units
of $\AA\ \mathrm{pixel^{-1}}$.}. The FWHM of the \ion{Mg}{2} emission is determined by a Gaussian fit 
through the SPLOT task. The measured two Lick indices, $D_{4000}$ and H$\delta_A$\footnote{
The index H$\delta_{\rm{A}}$ measures the EW of absorption in A type stars, and is defined
as by Worthey \& Ottaviani (1997) as 
$\mathrm{H}\delta_A=(4122.25-4083.50)(1-F_I/F_C)$
where $F_I$ is the flux within the $\lambda\lambda4083.50-4122.25$ feature bandpass, and $F_C$ the flux of
the pseudo-continuum within two defined bandpasses: blue $\lambda\lambda4041.60-4079.75$ and
red $\lambda\lambda4128.50-4161.00$.}, are listed 
in lines (11) and (12). The measured value of the Lick indices are in agreement with 
our spectral modelings described above.

\subsection{Eddington Ratio vs. $D_n(4000)$ Sequence}

The main goal of this paper is to study the coevolution of AGN and its host galaxy by taking 
advantage of the spectra of the hybrid QSOs.   
On account of the great progress in the reverberation mapping technique, 
a variety of empirical relationships are calibrated and used to estimate
the black hole viral masses ($M_{\mathrm{BH}}$) in AGNs (e.g., Kaspi et al. 2000, 2005; Peterson et al. 2004).
The commonly used calibrations are recently summarized in McGill et al. (2008).
The prominent broad \ion{Mg}{2} emission lines allow us to roughly estimate the 
$M_{\mathrm{BH}}$ and Eddington ratios ($L/L_{\mathrm{Edd}}$) for the three hybrid QSOs. 
\ion{Mg}{2} emission is an important coolant in high density BLR clouds in AGNs. Comparing with  
H$\beta$, the \ion{Mg}{2} emission is, in principle, less contaminated by star light. In addition, 
the equivalent width of \ion{Mg}{2} peaks at lower ionizing flux, which means
\ion{Mg}{2} is emitted from the region that has a larger distance from an isotropic 
ionizing source (e.g., Korista et al. 1997).

We estimate
the \ion{Mg}{2}-based $M_{\mathrm{BH}}$ according to the calibration
\begin{equation} 
M_{\mathrm{BH}}=2.04\bigg(\frac{L(3000\AA)}{10^{44}\ \mathrm{erg\ s^{-1}}}\bigg)^{0.88}
\bigg(\frac{\mathrm{FWHM(MgII)}}{\mathrm{km\ s^{-1}}}\bigg)^2M_\odot
\end{equation}
given in Kollmeier et al. (2006), where $L(3000\AA)$ is the AGN continuum 
luminosity at the rest-frame wavelength 3000\AA. The $L(3000\AA)$ is 
corrected for the intrinsic extinction according to the modeled color excess. 
The UV continuum-based calibration is 
adopted here because the total light 
spectra redward of the Balmer limit are dominated by the contribution from the 
starlight components.
The bolometric luminosities are then obtained from the estimated $L(3000\AA)$ by 
multiplying a factor of 5.9 (McLure \& Dunlop 2004). 
The estimated $M_{\mathrm{BH}}$ and $L/L_{\mathrm{Edd}}$ are listed in Table 2 for 
each object. 

The bolometric luminosities estimated from the modeled 
UV continuum are compared with that from the [\ion{O}{3}]$\lambda5007$ emission lines.
As a reasonable first approximation,  
[\ion{O}{3}] emission is believed to be isotropic in AGNs (Kuraszkiewicz et al. 2000). 
The isotropy of the [\ion{O}{3}] emission has been questioned by some studies of radio-loud AGNs 
(e.g., Baker \& Hunstead 1995; Jackson \& Rawlings 1997).
Despite the large scatter, the [\ion{O}{3}] luminosity ($L([\mathrm{OIII}])$) was reported 
to be correlated with the optical continuum luminosity for typical type I AGNs (e.g., Kauffmann et al. 2003a).
Given the relationship $L_{\mathrm{bol}}^{\mathrm{O3}}=3500L([\mathrm{OIII}])$,
the ratio $L^{\mathrm{UV}}_{\mathrm{bol}}/L_{\mathrm{bol}}^{\mathrm{O3}}=5.9L(3000\AA)/L_{\mathrm{bol}}^{\mathrm{O3}}$
is estimated to be 0.8 for SDSS\,J1634+4619, 0.5 for SDSS\,1026+1034, and 1.2 for SDSS\,0900+3813, which 
means a high consistence between the two independent estimations. 

As an additional test, the host stellar masses ($M_*$)
are estimated from the modeling of the observed spectra, and listed in
Table 2 as well. Similar as the results recently obtained in Alonso-Herrero et al. (2008),
the $M_*$ of the three hybrid QSOs are close to that of $z\approx2$ AGNs
($\sim10^{11}M_\odot$, e.g., Daddi et al. 2007; Kreik et al. 2007), and higher than those
of local AGNs ($\sim10^{10}M_\odot$, e.g., Kauffmann et al. 2003). The average ratio
$M_{\mathrm{BH}}/M_*$ is $\sim0.0019$ for the three hybrid QSOs, which is highly consistent with the
tight linear correlation between $M_{\mathrm{BH}}$ and the virial bulge mass.
The tight correlation established in the local Universe has an average ratio 
$\langle M_{\mathrm{BH}}/M_{\mathrm{Bulge}}\rangle \sim0.002$ (Marconi \& Hunt 2003).

The role of $L/L_{\mathrm{Edd}}$ in AGN evolution has been proposed for a long time (e.g., Grupe 2004; Mathur 2000). 
Wang \& Wei (2008) established a smooth $D_n(4000)-L/L_{\mathrm{Edd}}$ sequence by studying 
the nearby Seyfert 1.8/1.9 galaxies selected from the MPA/JHU SDSS DR4 catalog. 
The sequence indicates an evolutionary scenario that young AGN with
high $L/L_{\mathrm{Edd}}$ evolves to old AGN with low $L/L_{\mathrm{Edd}}$
along the sequence as the associated stellar population ages. Similar evolutionary sequences 
has been proposed by different authors through different methods and technologies (e.g., 
Wang et al. 2006; Kewley et al. 2006; Wild et al. 2007).  
As one generally believes that star formation activity decreases as stellar population ages,
the sequence is consistent with Watabe et al. (2008) who recently found
a close correlation between $L/L_{\mathrm{Edd}}$ and nuclear starburst luminosity assessed by the near
infrared PAH emission. Similar as done in our previous studies, 
$D_n(4000)$ is plotted against $L/L_{\mathrm{Edd}}$ by solid stars in Figure 4 for the three 
hybrid QSOs. The open circles show the $D_n(4000)-L/L_{\mathrm{Edd}}$
sequence that is established in Wang \& Wei (2008). The hybrid QSOs studied in this paper are 
clearly consistent with the $D_n(4000)-L/L_{\mathrm{Edd}}$ sequence reported in the local 
Seyfert galaxies, which implies that the evolution sequence could continue out to $z\approx0.5$. 

Note that the three hybrid QSOs are located in the middle range of the evolutionary sequence, 
which implies that 
they are at the transition stage not only from young to old AGNs, but also 
from host-dominated to AGN-dominated phase.
We propose that the hybrid QSOs are the progenitors of
local optical luminous QSOs. The three hybrid QSOs show recent starbursts within 1Gyr (recall 
that QSO UN\,J1025-0040 is associated with a 400Myr old post-starburst, Brotherton et al. 1999). On the 
contrary, relatively old (or old post-starburst) stellar populations are frequently identified in nearby luminous QSOs.
Nolan et al. (2001) found that the off-nuclear ($\approx$5\symbol{125}) stellar population is 
dominated by old stars ($\sim$8-14Gyr) for optically selected QSOs.   
Dunlop et al. (2003) observed a sample of local
QSOs at $z\sim0.2$ in imaging and spectroscopy. They claimed that the host galaxies of
these luminous QSOs are dominated by old stellar populations without recent massive star formation.
Canalizo et al. (2006) re-observed the 14 QSOs listed in the sample of Dunlop et al. (2003) by Keck LRIS
spectrograph. The spectra with high S/N ratios allow the authors to identify relatively old post-starbursts 
(0.6-2.2Gyr) in the host galaxies. These timescales are
comparable to the recent starburst ages recently inferred from the \it HST \rm ACS deep imaging study of
host galaxies of five low-redshift QSOs (Bennert et al. 2008). In addition, Tadhunter et al. (2005)
detected relatively old post-starbursts (0.1-2Gyr) in the off-nuclear regions of a few nearby
radio-loud AGNs.

Recent theoretical studies on the issue of co-evolution of AGN and its host galaxy suggest that 
AGNs are hard to be detected in the early host-dominated phase. Numerical simulations
of galaxies merger including SMBHs predict the theoretical light curves of the central AGN 
activity and associated star formation activity (e.g., Di Matteo et al. 2005; Springel et al. 2005).
At the beginning of evolution, the central AGN activity is predicted to be heavily obscured 
by the surrounding gas and dust, especially in UV/optical bands.  After the obscuration material 
is dispersed by the feedback from the accretion activity and emission from young, hot stars fades out
(i.e., at about 1Gyr after the beginning), luminous QSOs are observable because 
the star light from host is overwhelmed by the strong radiation from the luminous QSOs (Hopkins et al. 2005a,b).
An alternative possibility is the differential growth of the black hole mass and bulge mass (e.g., Weedman 1983).
This scenario is observationally supported by the detection of post-starburst stellar populations in 
Narrow-line Seyfert 1 galaxies with high $L/L_{\mathrm{Edd}}$ 
(e.g., Wang \& Wei 2006; Zhou et al. 2005). By observing local Seyfert galaxies 
with high spatial resolution down to 0.085\symbol{125}, Davies et al. (2007) recently
suggested that the black hole accretion delays for 50-100 Myr since the onset of star formation. 
The theoretical models developed by Kawakatu et al. (2003) and Granato et al. (2004)
predict that the change in phase from starburst-dominated to AGN-dominated takes place at 
a few $\times10^8$yr since the beginning of the star formation. 
Of course, we could not entirely exclude the possibility that the lack of extremely young
stellar population in our three hybrid QSOs is caused by the used $D_n(4000)$ criterion (i.e., $>1$). 

Our studies show that the hybrid QSOs could be an ideal laboratory for studying the co-evolution of 
AGN and its host galaxy because of their unique spectral properties. 
Additional deep spectroscopic
observations are required to search for more hybrid QSOs, and
to test the validity of the $D_n(4000)-L/L_{\mathrm{Edd}}$ sequence, especially in distant Universe.
Moreover, observations in infrared are helpful to constrain the dust content in these objects.

\subsection{Star Formation History/Rate}

There is accumulating evidence supporting that argument that star formation activity is suppressed in luminous AGNs 
(e.g., Ho 2005; Wang \& Wei 2008; Kim et al. 2006; Martin et al. 2007; Bundy et al. 2008;
Schawinski et al. 2007; Zheng et al. 2007). We argue that the current SFRs are significantly suppressed 
in the three hybrid QSOs. At first, the inserted panel in Figure 4 shows the $D_{4000}$ vs. H$\delta_\mathrm{A}$ 
diagram for the three hybrid QSOs. The dot-dashed line shows the stellar population
evolution locus for the model with exponentially decreasing SFR at solar metallicity 
($\psi(t)\propto e^{-t/(4\mathrm{Gyr})}$), and the dashed line the SSP model (BC03). The SSP model for 
a recent burst that ended 0.1-1 Gyr ago shows enhanced H$\delta_\mathrm{A}$ value because 
the optical spectrum is dominated by the emission of A-type stars. All three hybrid 
QSOs fall close to the single burst model due to their large H$\delta_\mathrm{A}$ values.  
Secondly, [\ion{O}{2}]$\lambda$3727 line emission is a good indicator of current SFR for 
starforming galaxies (e.g., Kennicutt 1998; Kewley et al. 2004). The spectra of both SDSS\,J1634+4619 
and SDSS\,J0900+3813 show marginally detectable [\ion{O}{2}]$\lambda$3727 emission features, which 
indicates that the current SFRs could be ignored in these two hybrid QSOs. [\ion{O}{2}]$\lambda$3727
emission is strong in SDSS\,J1026+1034. However, both AGN and \ion{H}{2} region can contribute to
[\ion{O}{2}] emission (e.g., Yan et al. 2006; Kim et al. 2006). The measured line ratio 
[\ion{O}{2}]/[\ion{O}{3}] is 0.54 (after the correction of the intrinsic extinction), and line ratio 
[\ion{O}{3}]/H$\beta_{\mathrm n}=4.0$. SDSS\,J1026+1034 
is therefore located in the region occupied by typical AGNs in the [\ion{O}{2}]/[\ion{O}{3}] vs.
[\ion{O}{3}]/H$\beta$ diagram (see Figure 7 in Kim et al. 2006). Moreover, the intrinsic 
[\ion{O}{3}] luminosity is estimated to be 
$L_{\mathrm{[OIII]}}\approx 2.0\times10^{42}\ \mathrm{erg\ s^{-1}}$.
At the given luminosity, the anticorrelation for typical AGNs 
between $L_{\mathrm{[OIII]}}$ and [\ion{O}{2}]/[\ion{O}{3}] 
predicts the line ratio [\ion{O}{2}]/[\ion{O}{3}] varies from 0.03 to 0.32 (see Figure 5 in Kim et al. 2006), which
means no more than 20\% of the total [\ion{O}{2}] emission is contributed from the star formation activity.
We consequently conclude that the narrow emission lines in SDSS\,J1026+1034 are mainly 
contributed from the central AGN.

\subsection{Radio Emission In SDSS\,J1634+4619}

The radio emission of SDSS\,J1634+4619 is detected by the NRAO VLA Sky Survey (NVSS; Condon et al. 1998) 
and Faint Images of the Radio Sky Survey at Twenty cm (FIRST) survey (Becker et al. 1995) at 
1.4GHz. The map of the FIRST survey shows an unresolved source with integrated flux $\sim4.30$mJy including 
a correction of 0.25mJy caused by Clean-Bias. The position of the radio source deviates from the corresponding 
optical source by 0.\symbol{125}36. The radio luminosity is calculated to be 
$P_{\mathrm{1.4GHz}}\simeq4.5\times10^{24}\ \mathrm{W\ Hz^{-1}}$ through a $k$-correction
by assuming a spectral shape $f_\nu\propto\nu^{-0.5}$ from optical to radio band.   

It is well known that the decimeter radio emission could be contributed from the supernova 
explosion of massive stars (i.e., $M\geq 8M_\odot$) with a life time $\approx 10^{7.5}$yr.  
In fact, the most radio-luminous starburst has a radio luminosity $\log(P_{\mathrm{5GHz}})\simeq22.3-23.4$
(Smith et al. 1998). The radio luminosity contributed from the recent starburst could be estimated from the 
past average SFR as $L_{1.4\mathrm{GHz}}=4.0\times10^{21} \mathrm{SFR}(\geq 5M_\odot)\ \mathrm{W\ Hz^{-1}}$ 
(Condon 1992). Recall that the observed spectrum has been modeled in terms of the linear combination 
of the five instantaneous bursts, 
we define the average SFR for each burst as $\mathrm{SFR}=\Delta M_*/\Delta t$, where $\Delta M_*$ is the star mass 
formed in each burst, and $\Delta t$ the step of time adopted in the spectral modeling. 
The average SFR of the latest burst is therefore 
estimated to be $\sim10^2 M_\odot\ \mathrm{yr^{-1}}$ for SDSS\,J1634+4619. 
Given the Salpeter IMF (Salpeter 1955), the starburst-contributed radio luminosity   
is $\sim8\times10^{22}\ \mathrm{W\ Hz^{-1}}$, which is one order of magnitude 
lower than the calculated total radio luminosity.

\subsection{A Companion Galaxy Of SDSS\,J1634+4619?}

It is widely accepted that galaxies interaction can trigger extreme nuclear accretion and 
star formation activity (e.g., Toomre \& Toomre 1972; Larson \& Tinsely 1978; Sanders et al. 1988; 
Stockton 1999; Heckman et al. 1984; Di Matteo et al. 2005; Springel et al. 2005a, b). 
Recent \it HST \rm ACS deep imaging study identified 
significant fine structures such as shells and tidal tails in a small sample of host galaxies of 
low redshift QSOs (Bennert et al. 2008).
A companion galaxy at post-starburst phase ($\sim800$Myr) is identified in the  
particular case UN\,J1025-0040 
(Brotherton et al. 1999; Canalizo et al. 2000). The \it HST \rm image shows an interacting companion
for the ``Q+A'' object SDSS\,J231055-090107 (Canalizo et al. 2006).

The optical images of SDSS\,J1634+4619 taken by SDSS show an unresolved source 
neighbored by a marginally detectable, very faint source to southwest in appearance. 
Figure 5 shows the $r'$-band image. 
The intensity contours are over-plotted on 
the image after the background level determined around the source is subtracted.  
To identify the faint companion more clearly, we project the image along the long sides of 
the two rectangles 
overlaid in Figure 5. Figure 6 presents the count distributions along the two
directions. As shown by the left panel in Figure 6, the peaks of SDSS\,J1634+4619 and 
the possible companion are separated by about 8 pixels. This separation corresponds to a projected
physical distance $\sim$11\,kpc if the companion is at z=0.576. 
Finally, more deep spectroscopic observations and imagings are
necessary to determine whether the companion interacts with SDSS\,J1634+4619 physically 
or is just a foreground star/galaxy.  

\section{SUMMARY}

We select 15 intermediate-z galaxies with hybrid spectra from SDSS DR6 to 
study the issue of AGN-host connection. The spectra redward of the Balmer limit
are dominated by starlight, and the spectra at blue end by both an AGN continuum and a
\ion{Mg}{2} broad emission line. The spectra are analyzed in detail in three objects:
SDSS\,J162446.49+461946.7, SDSS\,J102633.32+103443.8 and SDSS\,J090036.44+381353.0.
Without intensive current star formation activities, the modeled recent burst ages
range from $\sim$400Myr to 1Gyr. Based on the \ion{Mg}{2}-based black hole masses,
the three hybrid QSOs are consistent with the $D_n(4000)-L/L_{\mathrm{Edd}}$ sequence previously
established in local AGNs.

\acknowledgments
We would like to thank the anonymous referee for very useful comments and important 
suggestions that improved the presentation.
This search has made use of the NASA/IPAC Extragalactic Database, which is 
operated by JPL, Caltech, under contract with the NASA. The SDSS archive 
data are created and distributed by the Alfred P. Sloan Foundation. This work
is supported by the National Science Foundation of China (under grants 10503005 and 10803008),
and by the National Basic Research Program of China (grant 2009CB824800).

\clearpage



\begin{figure}
\epsscale{.80}
\plotone{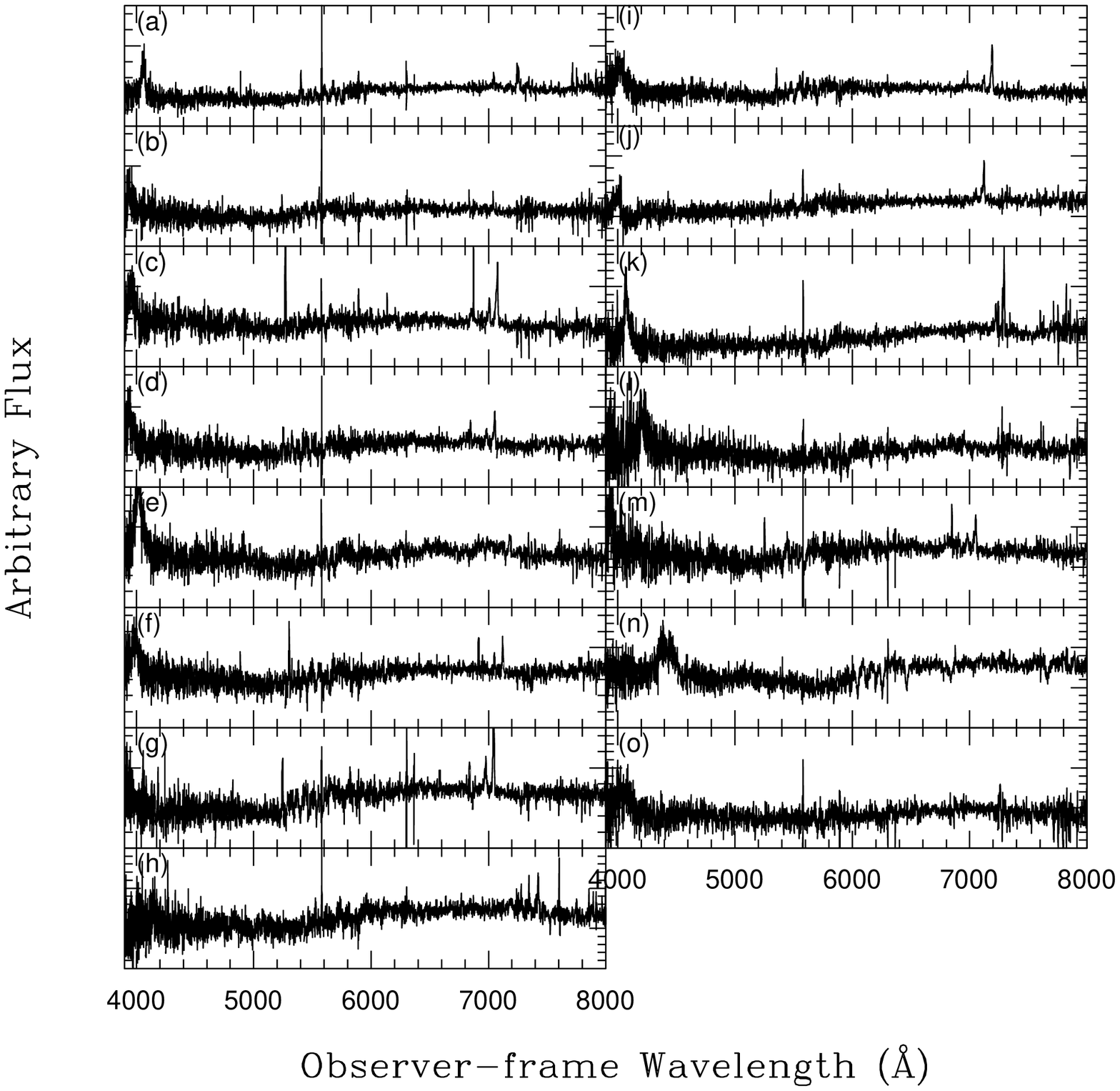}
\caption{Observed spectroscopy atlas for the 15 hybrid QSOs selected in this paper. (a) SDSS\,J074156.79+345405.5. 
(b) SDSS\,J074524.97+375436.7. (c) SDSS\,J081535.89+552558.4. (d) SDSS\,J082718.94+294204.3. 
(e) SDSS\,J090036.44+381353.0. (f) SDSS\,J093912.82+455358.8. (g) SDSS\,J101011.59+444212.0.
(h) SDSS\,J101036.80+294520.1. (i) SDSS\,J102633.32+103443.8. (j) SDSS\,J115507.20+351058.7. 
(k) SDSS\,J141324.27+530527.0. (l) SDSS\,J154901.16+071247.6. (m) SDSS\,J160616.23+223242.1.
(n) SDSS\,J163446.49+461946.7. (o) SDSS\,J225106.81-080107.8.}
\end{figure}

\begin{figure}
\epsscale{.80}
\plotone{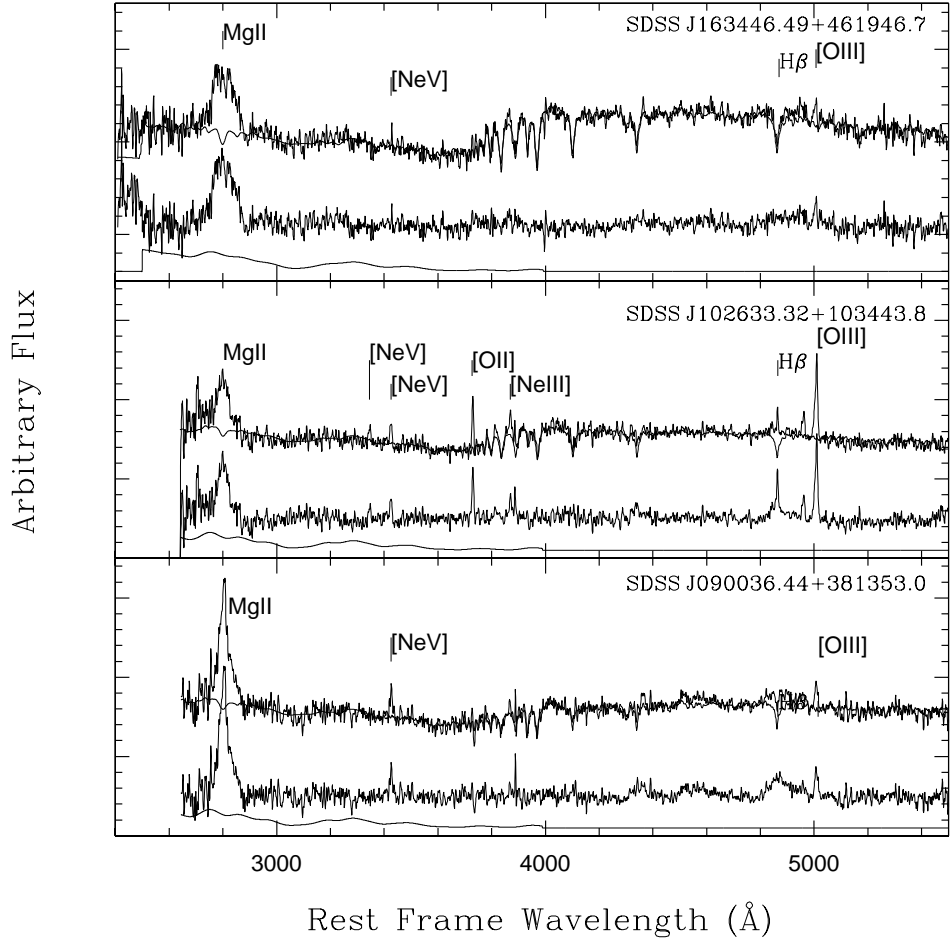}
\caption{An illustration of the spectral modeling for the starlight templates with 
the solar metallicity. In each panel, the observed spectrum is shown by the top curve. 
The modeled spectrum is over-plotted on the observed spectrum. The middle and 
bottom curves shows the residual emission-line spectrum and modeled 
UV \ion{Fe}{2} complex.} 
\end{figure}

\begin{figure}
\epsscale{.80}
\plotone{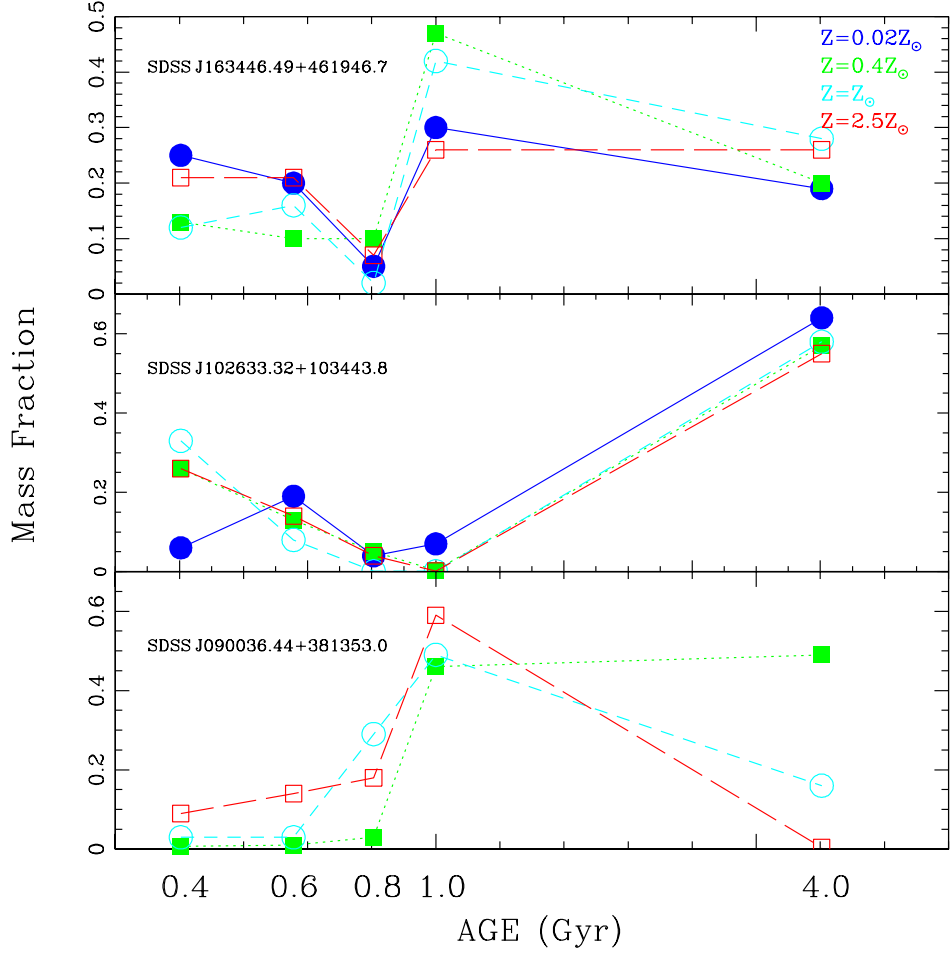}
\caption{The modeled mass fractions of the adopted five/four instantaneous bursts with different ages. 
The ages of the bursts
range from 0.4 to 4.0Gyr. The results with different metallicities are shown by different 
colors, point/line types for clarification.
}
\end{figure}

\begin{figure}
\epsscale{.80}
\plotone{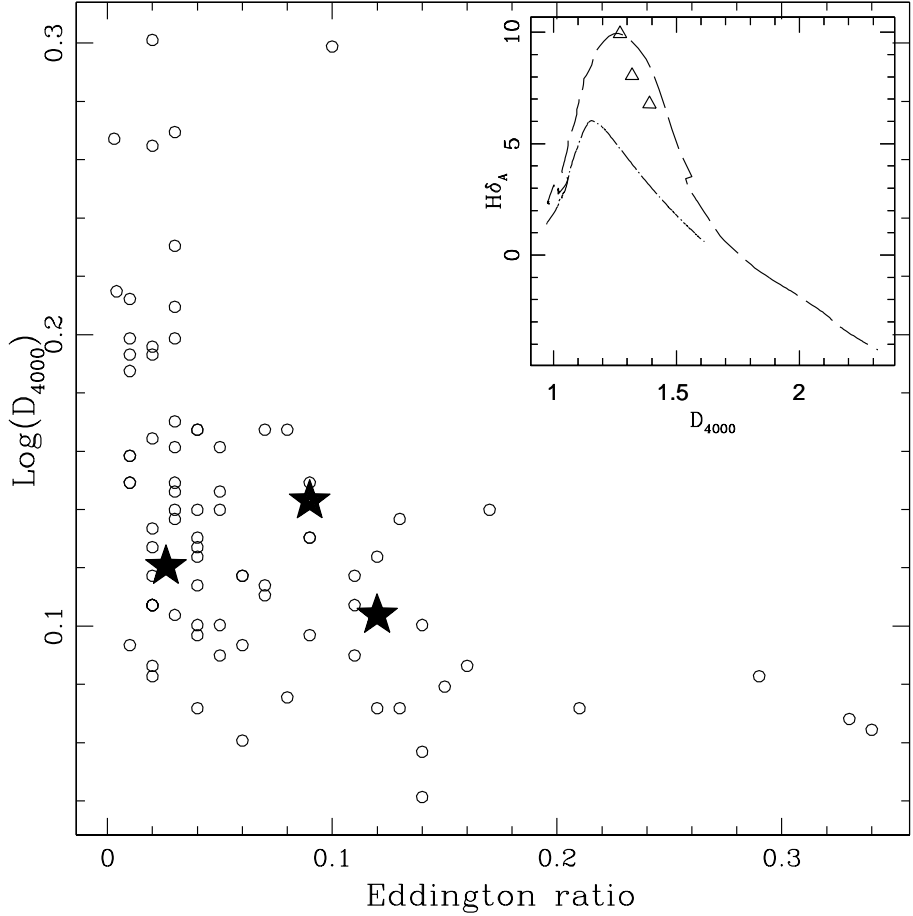}
\caption{$D_n(4000)$ plotted against $L/L_{\mathrm{Edd}}$. The local objects studied in 
Wang \& Wei (2008) are shown by the open 
circles. The solid stars mark the positions of the three hybrid QSOs studied in this paper.
Their black hole masses are estimated from the \ion{Mg}{2} broad emission lines. \it Inserted
panel: \rm the $D_{4000}-\mathrm{H\delta_A}$ diagram for the three hybrid QSOs (open triangles). 
The dashed line shows the stellar population evolution locus of the SSP model with the 
solar metallicity, and 
the dot-dashed line the model with exponentially decreasing star formation rate 
$\psi(t)\propto e^{-t/(\mathrm{4Gyr})}$.  
}
\end{figure}

\begin{figure}
\epsscale{.80}
\plotone{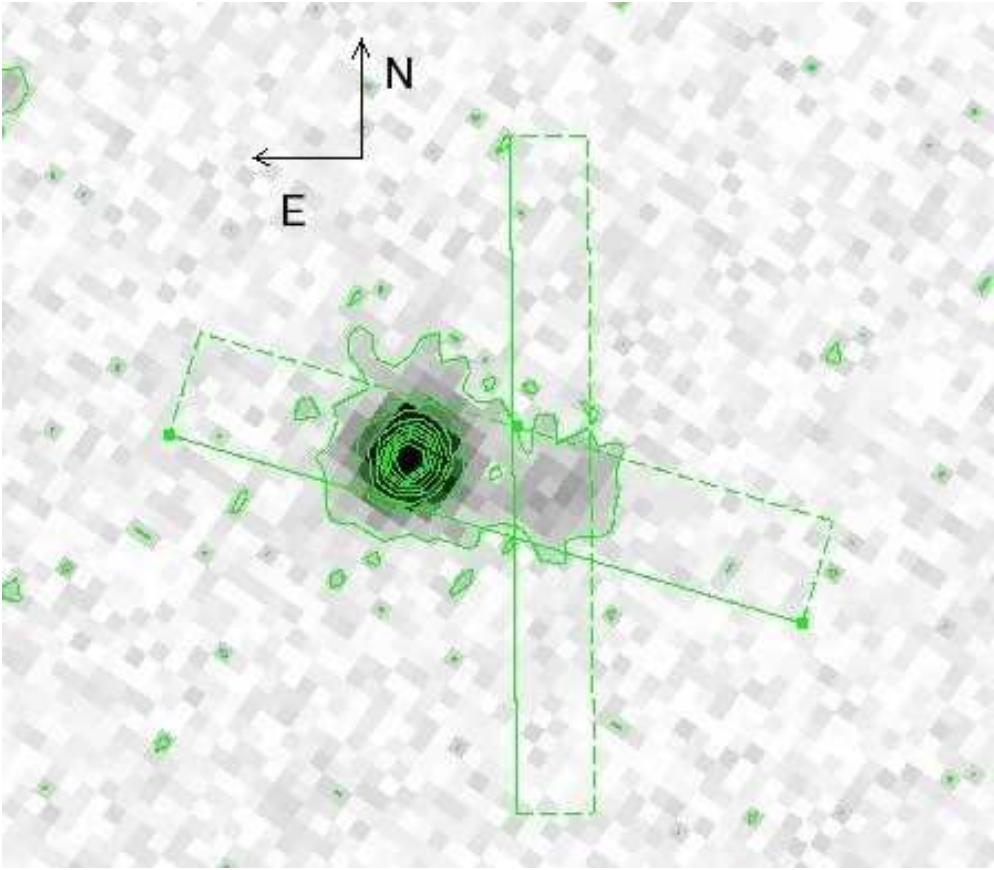}
\caption{SDSS $r'$-band image of SDSS\,J1634+4629 after subtract the background level determined around the object. 
North and east are indicated by the arrows shown at left-top corner. The intensity contours are overlaid by 
the solid curves. The rectangles shown by the dashed lines mark the directions (the long side)
used to project the map. }
\end{figure}

\begin{figure}
\epsscale{.80}
\plotone{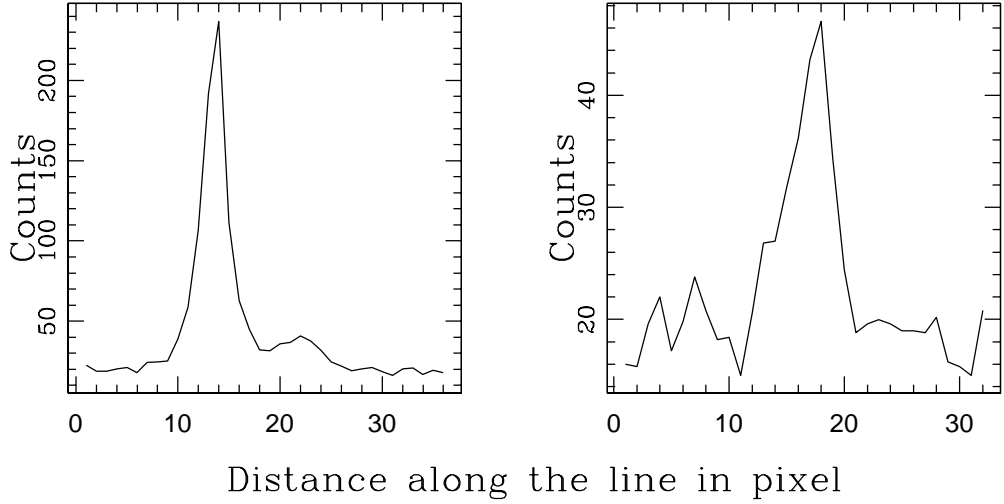}
\caption{Projections on the two directions shown in Figure 5. The horizontal axis is the distance along the 
direction in units of pixel.}
\end{figure}

\begin{table}
\begin{center}
\caption{THE SAMPLE OF 15 HYBRID QSOs SELECTED FROM SDSS DR6}
\scriptsize
\begin{tabular}{cccccccc}
\tableline\tableline
Name & fiber ID & Plate ID & MJD ID & z & EW(\ion{Mg}{2})\tablenotemark{\dagger} & S/N \ion{Mg}{2}\tablenotemark{\dagger} & S/N spec\tablenotemark{\ddagger}\\
     &          &          &        &   &   \AA           &    & \\        
(1) & (2) & (3) & (4) & (5) & (6) & (7) & (8)\\
\tableline
SDSS\,J074156.79+345405.5 & 394 & 542 & 51993 & 0.448 & $32.3\pm1.7$ & 18.9 & 5.0\\
SDSS\,J074524.97+375436.7 & 170 & 433 & 51873 & 0.406 & $15.2\pm1.7$ & 8.7 & 5.0\\
SDSS\,J081535.89+552558.4 & 590 & 1871 & 53384 & 0.413 & $16.9\pm2.3$ & 7.3 & 4.7\\
SDSS\,J082718.94+294204.3 & 600 & 1207 & 52672 & 0.408 & $19.8\pm2.1$ & 9.3 & 5.5\\
SDSS\,J090036.44+381353.0\tablenotemark{*} & 478 & 936 & 52705 & 0.434 & $41.9\pm2.0$ & 21.2 & 5.9\\
SDSS\,J093912.82+455358.8 & 380 & 1202 & 52672 & 0.422 & $21.4\pm1.9$ & 11.5 & 6.2\\
SDSS\,J101011.59+444212.0 & 193 & 943 & 52376 & 0.406 & $4.7\pm1.3$ & 3.6 & 3.9\\
SDSS\,J101036.80+294520.1 & 69 & 1953 & 53358 & 0.483 & $7.3\pm2.1$ & 3.5 & 5.2\\
SDSS\,J102633.32+103443.8\tablenotemark{*} & 142 & 1598 & 53033 & 0.435 & $23.4\pm2.2$ & 10.8 & 6.0\\
SDSS\,J115507.20+351058.7 & 375 & 2099 & 53469 & 0.421 & $371.1\pm21.8$ & 17.0 & 2.1\\
SDSS\,J141324.27+530527.0 & 133 & 1325 & 52762 & 0.455 & $56.8\pm3.3$ & 17.1 & 2.9\\
SDSS\,J154901.16+071247.6 & 272 & 1727 & 53859 & 0.501 & $27.4\pm2.5$ & 10.8 & 4.2\\
SDSS\,J160616.23+223242.1 & 237 & 1852 & 53534 & 0.408 & $23.7\pm2.7$ & 8.8 & 5.6\\
SDSS\,J163446.49+461946.7\tablenotemark{*} & 163 & 627 & 52144 & 0.576 & $10.8\pm1.1$ & 10.0 & 5.3\\
\tableline
\end{tabular}
\tablenotetext{*}{Studied in this paper.}
\tablenotetext{\dagger}{The signal-to-noise ratio of \ion{Mg}{2} emission line measured by the SDSS pipelines.}
\tablenotetext{\ddagger}{The average signal-to-noise ratio of whole observed spectrum}
\end{center}
\end{table}

\begin{table}
\begin{center}
\caption{PROPERTIES OF THE THREE \ion{Mg}{2} BROAD EMISSION LINE
SELECTED HYBRID QSOs}
\begin{tabular}{lccc}
\tableline\tableline
Properties & SDSS\,J1634+4619 & SDSS\,J1026+1034 & SDSS\,0900+3813\\
(1) & (2) & (3) & (4)\\
\tableline
z\dotfill(1) & 0.576 & 0.435 & 0.434 \\
E(B-V)\dotfill(2) & 0.11 & 0.26 & 0.22\\
$F_{\lambda,3000\AA}/10^{-16}\ \mathrm{ergs\ s^{-1}\ cm^{-2}\ \AA^{-1}}$\ (3) &  $1.12\pm0.13$ & $2.58\pm0.23$ & $2.26\pm0.32$\\
$F_{\mathrm{[OIII]}}/10^{-16}\ \mathrm{ergs\ s^{-1}\ cm^{-2}}$\dotfill(4)  &  $4.85\pm1.10$ & $28.3\pm6.4$ & $9.77\pm0.08$\\
$F_{\mathrm{[OII]}}/10^{-16}\ \mathrm{ergs\ s^{-1}\ cm^{-2}}$\dotfill(5)  & \dotfill & $15.4\pm3.8$ & \dotfill\\   
FWHM(\ion{Mg}{2})/$\mathrm{km\ s^{-1}}$\dotfill(6) &  $9300\pm500$ & $4500\pm300$  & $5000\pm800$\\
$M_{\mathrm{BH}}/M_\odot$\dotfill(7) &  $6.8\times10^{8}$ & $1.8\times10^{8}$ & $1.9\times10^{8}$\\
$L/L_{\mathrm{Edd}}$\dotfill(8) &  0.026 & 0.12 & 0.09\\
$M_*/M_\odot$\dotfill(9) & $2.4\times10^{11}$ & $1.2\times10^{11}$ & $1.4\times10^{11}$\\
S.P. Age/Gyr\dotfill(10) &   $\sim0.6$ &  $\leq0.4$ & $\sim0.8-1.0$\\ 
$D_n(4000)$\dotfill(11) &  1.32 & 1.27 &  1.39\\
H$\delta_A$\dotfill(12) &  8.06 & 9.93 & 6.78\\
\tableline
\end{tabular}
\end{center}
\end{table}







\begin{thebibliography}{}
\bibitem[Adelman-McCarthy et al. 2008]{ald08} Adelman-McCarthy, J. K., et al. 2008, \apjs, 175, 297
\bibitem[Alonso-Herrero et al. 2008]{alo08} Alonso-Herrero, A., Perez-Gonzelez, P. G., Rieke, G. H., et al. 2008, \apj, 677, 127
\bibitem[Antonucci 1993]{ant93} Antonucci, R. R. J. 1993, \araa, 31, 473
\bibitem[Baker \& Hunstead 1995]{bak95} Baker, J. C., \& Hunstead, R. W. 1995, \apj, 452, 95 
\bibitem[Balogh et al. 1999]{bal99} Balogh, M. L., et al. 1999, \apj, 527, 54
\bibitem[Becker et al. 1995]{bec95} Becker, R. H., White, R. L., \& Helfand, D. J. 1995, \apj, 450, 559
\bibitem[Bennert et al. 2008]{ben08} Bennert, N., Canalizo, G., Jungwiert, B., et al. 2008, \apj, 677, 846
\bibitem[Begelman \& Nath 2005]{beg05} Begelman, M. C., \& Nath, B. B. 2005, \mnras, 361, 1387
\bibitem[Bruhweiler \& Verner 2008]{bru08} Bruhweiler, F., \& Verner, E. 2008, 675, 83 
\bibitem[Bromley et al. 1998]{bro98} Bromley, B. C., et al. 1998, \apj, 505, 2
\bibitem[Brotherton et al. 1999]{bro00} Brotherton, M. S., van Breugel, Wil., Stanford, S. A., et al. 1999, \apjl, 520, 87
\bibitem[Bruzual 1983]{bru83} Bruzual, A. G. 1983, \apj, 273, 105
\bibitem[Bruzual \& Charlot 2003]{bc03} Bruzual. G., \& Charlot. S. 2003, \mnras, 344, 1000
\bibitem[Bundy et al. 2005]{bud05} Bundy, K., Ellis, R. S., \& Conselice, C. J. 2005, \apj, 625, 621 
\bibitem[Canalizo et al. 2000]{can00} Canalizo, G., Stockton, A., Brotherton, M. S., et al. 2000, \aj, 119, 59
\bibitem[Canalizo et al. 2006]{can06} Canalizo, G., Stockton, A., Brotherton, M. S., et al. 2006, NewA, 2006, 50, 650 
\bibitem[Cardilli et al. 1989]{car89} Cardilli, J. A., Clayton, G. C., \& Mathis, J. S. 1989, \apj, 345, 245
\bibitem[Condon 1992]{con92} Condon, J. J. 1992, \araa, 30, 575
\bibitem[Condon et al. 1998]{con98} Condon, J. J., Cotton, W. D., Greisen, E. W., et al. 1998, \aj, 115, 1639
\bibitem[Daddi et al. 2007]{dad07} Daddi, E., et al. 2007, \apj, 670, 173
\bibitem[Davies et al. 2007]{dav07} Davies, R. I., Mueller Sanchez, F., Genzel, R., et al. 2007, \apj, 671, 1388
\bibitem[Di Matteo et al. 2005]{dim05} Di Matteo, T., Springel, V., \& Hernquist, L. 2005, Nature, 433, 604
\bibitem[Dunlop et al. 2003]{dun03} Dunlop, J. S., McLure, R. J., Kukula,M. J., et al. 2003, \mnras, 340, 1095
\bibitem[Elitzur 2007]{eli07} Elitzur, M. 2007, ASPC, 373, 415
\bibitem[Fabian 1999]{fab99} Fabian, A. C. 1999, \mnras, 308, L39 
\bibitem[Ferrarese \& Merritt 2000]{fem00} Ferrarese, L., \& Merritt, D. 2000, \apjl, 529, 13
\bibitem[Ferrarese et al. 2006]{fer06} Ferrarese, L., Cote, P., Dalla Bonta, E., et al. 2006, \apj, 644, L21
\bibitem[Gebhardt et al. 2000]{ge00} Gebhardt, K., Bender, R., Bower, G., et al. 2000, \apjl, 539, 13
\bibitem[Granato et al. 2004]{gra04} Granato, J. E., et al. 2004, \apj, 600, 580
\bibitem[Glazebrook et al. 1998]{gla98} Glazebrook, K., Offer, A., R., \& Deeley, K. 1998, \apj, 492, 98
\bibitem[Greene \& Ho (2006)]{gh06} Greene, J. E., \& Ho, L. C. 2006, \apjl, 641, L21
\bibitem[Heckman et al. 1984]{hec84}Heckman, T. M., Bothun, G. D., Balick, B., et al. 1984, \apj, 89, 7
\bibitem[Heckman et al. 2004]{he04} Heckman, T. M., Kauffmann, G., Brinchmann, J., et al. 2004. \apj, 613, 109
\bibitem[Ho 2005]{ho05} Ho, L. C. 2005, \apj, 629, 680
\bibitem[Ho et al. 2008a]{ho08a} Ho, L. C., Darling, J., \& Greene, J. E. 2008a, \apj, 681, 128
\bibitem[Ho et al. 2008b]{ho08b} Ho, L. C., Darling, J., \& Greene, J. E. 2008b, \apjs, 177, 103 
\bibitem[Hopkins et al. 2005a]{hop05a} Hopkins, P. F., Hernquist, L., Cox, T. J., et al. 2005a, \apj, 630, 705
\bibitem[Hopkins et al. 2005b]{hop05b} Hopkins, P. F., Hernquist, L., Martini, P., et al. 2005b, \apjl, 625, 71
\bibitem[Hopkins et al. 2006]{hop06} Hopkins, P. F., Hernquist, L., Cox, T. J., et al. 2006, \apjs, 163, 1
\bibitem[Jackson \& Rawlings 1997]{jac97} Jackson, N., \& Rawlings, S. 1997, \mnras, 286, 241
\bibitem[Kaspi et al. 2005]{kas05} Kaspi, S., Maoz, D., Netzer, H., et al. 2005, \apj, 629, 61
\bibitem[Kaspi et al. 2000]{kas00} Kaspi, S., Smith, P. S., Netzer, H., et al. 2000, \apj, 533, 631
\bibitem[Kauffmann et al. 2003a]{ka03a}Kauffmann, G., Heckman, T. M., Tremonti, C., et al. 2003a, \mnras, 346, 1055
\bibitem[Kauffmann et al. 2003b]{ka03b}Kauffmann, G., Heckman, T. M., White, S. D. M., et al. 2003b, \mnras, 341, 54
\bibitem[Kauffmann et al. 2003c]{ka03c}Kauffmann, G., Heckman, T. M., White, S. D. M., et al. 2003c, \mnras, 341, 33
\bibitem[Kewley et al. 2004]{kew04} Kewley, L. J., Geller, M. J., \& Jansen, R. A. 2004, \aj, 127, 2002
\bibitem[Kewley et al. 2006]{kew06} Kewley, L. J., Groves, B., Kauffmann, G., \& Heckman, T. 2006, \mnras, 372, 961
\bibitem[Kim et al. 2006]{kim06} Kim, M., Ho, L. C., \& Im, M. 2006, \apj, 642, 702
\bibitem[Kollmeier et al. 2006]{kol06} Kollmeier, J. A., Onken, C. A., Kochanek, C. S., et al. 2006, \apj, 648, 128
\bibitem[Korista et al. 1997]{kor97} Korista, K., Baldwin, J., Ferland, G., et al. 1992, \apjs, 108, 401
\bibitem[Kuraszkiewicz et al. 2000]{kur00} Kuraszkiewicz, J., Wilkes, B. J., Brandt, W. N., et al. 2000, \apj, 542, 631
\bibitem[Larson \& Tinsley 1978]{lar78} Larson, R. B., \& Tinsley, B. M. 1978, \apj, 219, 46 
\bibitem[Magrorrian et al. 1999]{mar99} Magorrian, J., et al. 1999, \aj, 115, 2285
\bibitem[Marconi \& Hunt 2003]{mar03} Marconi, A., \& Hunt, L. K. 2003, \apj, 589, L21
\bibitem[Martin et al., 2007]{mar07} Martin, D. C., et al. 2007, \apjs, 173, 342
\bibitem[Mathur 2000]{mar00} Mathur, S. 2000, \mnras, 314, L17
\bibitem[McGill et al. 2008]{mcg08} McGill, K. L., Woo, Jong-Hak., Treu, T., et al. 2008, \apj, 673, 703
\bibitem[McLure \& Dunlop 2004]{mcl04} McLure, R. J., \&  Jarvis, M. J. 2004, \mnras, 353, L45
\bibitem[Nandra et al. 2005]{nan05} Nandra, K., Laird, E. S., \& Steidel, C. C. 2005, \mnras, 360, L39
\bibitem[Nolan et al. 2001]{nol01} Nolan, L. A., Dunlop, J. S., Kukula, M. J., et al. 2001, \mnras, 323, 308
\bibitem[Perez-Montero \& Diaz 2003]{per03} Perez-Montero, E., \& Diaz, A. I. 2003, \mnras, 346, 105
\bibitem[Peterson et al. 2004]{pet04} Peterson, B. M., et al. 2004, \apj, 613, 682
\bibitem[Salpeter 1955]{sal55} Salpeter, E. E. 1955, \apj, 121, 161
\bibitem[Sanders et al. 1988]{san88} Sanders, D. B., Soifer, B. T., Elias, J. H., et al. 1988, \apj, 325, 74
\bibitem[Schawinski et al. 2007]{sch07} Schawinski, K., Thomas, D., Sarzi, M., et al. 2007, \mnras, 382, 1415 
\bibitem[Schlegel et al. 1998]{sch98} Schlegel, D, Finkbeiner, D. P., \& Davis, M. 1998, \apj, 500, 525
\bibitem[Smith et al. 1998]{smi98} Smith, D. A., Herter, T., \& Haynes, M. P. 1998, \apj, 494, 150
\bibitem[Spergel et al. 2003]{spe03} Spergel, D. N., et al. 2003, \apjs, 148, 175
\bibitem[Springel et al. 2005a]{spr05a} Springel, V., Di Matteo, T., \& Hernquist, L. 2005a, \apj, 620, L79 
\bibitem[Springel et al. 2005b]{spr05b} Springel, V., Di Matteo, T., \& Hernquist, L. 2005a, \mnras, 339, 289 
\bibitem[Stockton 1999]{sto99} Stockton, A. 1999, IAUS, 186, 311
\bibitem[Tadhunter et al. 2005]{tad05} Tadhunter, C., Robinson, T. G., Gonzalez Delgado, R. M., et al. 2005, \mnras, 356, 480
\bibitem[Toomre \& Toomre 1972]{too72} Toomre, A., \& Toomre, J. 1972, \apj, 178, 623
\bibitem[Tremaine et al. 2002]{tre02} Tremaine, S., Gebhardt, K., Bender, R., et al. 2002, \apj, 574, 740
\bibitem[Wang \& Wei 2006]{wan06} Wang, J., \& Wei, J. Y. 2006, \apj, 648, 158
\bibitem[Wang \& Wei 2008]{wan08} Wang, J., \& Wei, J. Y. 2008, \apj, 679, 86
\bibitem[Watabe et al. 2008]{wat08} Watabe, Y., Kawakatu, N., \& Imanishi, M. 2008, \apj, 677, 895 
\bibitem[Wild et al. 2007]{wil07} Wild, V., et al. 2007, \mnras, 381, 543
\bibitem[Worthey \& Ottaviani 1997]{wo97} Worthey, G., \& Ottaviani, D. L. 1997, \apjs, 111, 377 
\bibitem[York et al. 2000]{yo00} York, D. G., et al. 2000, \aj, 120, 1579
\bibitem[Zheng et al. 2007]{zhe07} Zheng, X. Z., et al. 2007, \apj, 661, L41
\bibitem[Zhou et al. 2005]{zho05} Zhou, H. Y., Wang, T. G., Dong, X. B., et al. 2005, Mem. Soc. Astron. Italiana, 76, 93
\end{thebibliography}
\end{document}